# IC-U-Net: A U-Net-based Denoising Autoencoder Using Mixtures of Independent Components for Automatic EEG Artifact Removal

Chun-Hsiang Chuang[a,b], Kong-Yi Chang[c], Chih-Sheng Huang[d], Tzyy-Ping Jung[e]

[a] Research Center for Education and Mind Sciences, College of Education, National Tsing Hua University, Hsinchu, Taiwan

[b] Institute of Information Systems and Applications, College of Electrical Engineering and Computer Science, National Tsing Hua University, Hsinchu, Taiwan

[c] Department of Computer Science and Engineering, National Taiwan Ocean University, Keelung, Taiwan.

[d] Department of Artificial Intelligence Research and Development, Elan Microelectronics Corporation, Hsinchu, Taiwan

[e] Institute of Engineering in Medicine and Institute for Neural Computation, University of California, San Diego, La Jolla, USA

* Correspondence should be addressed to either of the following:
Chun-Hsiang Chuang,
Center for Education and Mind Sciences, College of Education,
Institute of Information Systems and Applications, College of Electrical Engineering and Computer Science,
National Tsing Hua University, Taiwan
Phone: +886 (03) 5715131#78607
E-mail: ch.chuang@mx.nthu.edu.tw

**Abstract**

Electroencephalography (EEG) signals are often contaminated with artifacts. It is imperative to develop a practical and reliable artifact removal method to prevent misinterpretations of neural signals and underperformance of brain-computer interfaces. This study developed a new artifact removal method, IC-U-Net, which is based on the U-Net architecture for removing pervasive EEG artifacts and reconstructing brain sources. The IC-U-Net was trained using mixtures of brain and non-brain sources decomposed by independent component analysis and employed an ensemble of loss functions to model complex signal fluctuations in EEG recordings. The effectiveness of the proposed method in recovering brain sources and removing various artifacts (e.g., eye blinks/movements, muscle activities, and line/channel noises) was demonstrated in a simulation study and three real-world EEG datasets collected at rest and while driving and walking. IC-U-Net is user-friendly and publicly available, does not require parameter tuning or artifact type designations, and has no limitations on channel numbers. Given the increasing need to image natural brain dynamics in a mobile setting, IC-U-Net offers a promising end-to-end solution for automatically removing artifacts from EEG recordings.

# 1. Introduction

Electroencephalography (EEG), a non-invasive and convenient modality for measuring high-temporal-resolution changes in electrical potentials over the human scalp, is one of the dominant neuroimaging modalities used in brain–computer interfaces (BCIs) (Gu *et al.*, 2021; Lazarou *et al.*, 2018; Nicolas-Alonso and Gomez-Gil, 2012; Rashid *et al.*, 2020). Advances in signal processing and analytic methods have expanded the capabilities and applicability of EEG in real-world practices. Nevertheless, there is an emerging need to measure brain dynamics in moving humans (Jungnickel and Gramann, 2016), and one of the associated challenging issues is removing artifacts from EEG signals (Gwin *et al.*, 2010; Jiang *et al.*, 2019) with minimum distortion and contamination. A practical and effective method for suppressing artifacts and reconstructing EEG signals would considerably improve the signal-to-noise ratio (SNR) of the recordings, enhance the performance of BCIs, and increase the accuracy of data interpretation (Haumann *et al.*, 2016; Radüntz *et al.*, 2015).

The realization of such a technology has taken center stage in the research on EEG signal processing. Several researchers have proposed various countermeasures to remove signal contamination caused by, for example, eye blinks/movements, muscle activity, and line noise. One commonly adopted countermeasure is blind source separation (BSS) (Fitzgibbon *et al.*, 2007; Joyce *et al.*, 2004; Jung *et al.*, 2000), in which various statistical assumptions, such as independence (Jung *et al.*, 2000; Makeig *et al.*, 1996), correlation (de Cheveigné *et al.*, 2018; Lin *et al.*, 2018), covariance (Chang *et al.*, 2020), or stationarity (Maddirala and Veluvolu, 2021; von Bunau *et al.*, 2010), are used to extract the underlying sources from their mixtures at sensors. Artifact subspace reconstruction (ASR) (Chang *et al.*, 2020), for instance, assumes that the EEG data share the same principal component space, projects raw EEG signals onto this feature space, and then reconstructs signals by preserving those projected principal components whose variance does not exceed a certain predefined threshold. Another well-grounded and effective BSS method is independent component analysis (ICA) (Jung *et al.*, 2000; Makeig *et al.*, 1996), which separates instantaneously and linearly mixed recordings into mutually independent components (ICs). Given that an IC classifier (Pion-Tonachini *et al.*, 2019; Winkler *et al.*, 2011) can automatically differentiate artifactual ICs from brain ICs, artifact-corrected EEG signals can be obtained by pruning non-brain sources and backprojecting brain sources to the sensor level. Numerous studies, including those discussed herein, have advocated for better EEG representation through BSS, which would allow deeper scientific analysis and broader use of EEG data. However, most BSS techniques seek a single linear or nonlinear mapping function between source signals and sensor observations, which might be too simplistic for the complex nature of brain activity.

Recently, deep learning models have been widely applied to EEG analysis and BCIs (Lawhern *et al.*, 2018; Schirrmeister *et al.*, 2017), with the key idea being to extract the underlying feature representation (Bengio *et al.*, 2013; Shorten and Khoshgoftaar, 2019) using a powerful neural network. Unlike conventional BSS methods, deep learning enables the construction of data representations without the need of restrictive assumptions. One such representative feature learning model is the autoencoder (Hinton and Salakhutdinov, 2006). For example, the denoising autoencoder (DAE) (Vincent *et al.*, 2010) has great potential for removing EEG artifacts (Leite *et al.*, 2018), as evidenced by its effectiveness in other applications, such as removing noise from images (Phadikar *et al.*, 2020), restoring speech (Lu *et al.*, 2013), decoding incomplete EEG recordings (Li *et al.*, 2015), and enhancing ECG signals (Chiang *et al.*, 2019; Xiong *et al.*, 2016). Another representative encoder–decoder scheme is U-Net (Ronneberger *et al.*, 2015) that originated from convolutional neural networks and was initially developed for biomedical image segmentation. U-Net and its variants have recently been successfully applied to image denoising (Guan *et al.*, 2020; Gurrola-Ramos *et al.*, 2021) and various EEG applications (Chatzichristos *et al.*, 2020; Perslev *et al.*, 2019; You *et al.*, 2021). However, to the best of our knowledge, no study has yet used U-Net to denoise EEG signals. Because encoder and decoder networks can serve as the source separation and mixing, respectively (Fig. 1), we conjecture that U-Net can recover brain sources and thereby suppress artifacts. Furthermore, because of its simple structure and its minimal training sample requirement, U-Net is well suited for processing EEG signals, which are typically difficult to accumulate on a large scale.

In artifact removal, the availability of noisy–free EEG pairs is crucial and considerably affects the signal reconstruction performance regardless of the DAE model used. Such pairs can be obtained by backprojecting selected brain sources decomposed by BSS to the sensor level, in addition to generating and adding artificial noises to EEG data (Delorme *et al.*, 2007; Li *et al.*, 2021; Val-Calvo *et al.*, 2019; Zeng *et al.*, 2016). For instance, the spatial mixing matrix of ICA can be used to thoroughly mix brain and non-brain sources to generate clean EEG targets and noisy EEG inputs. Each spatial mixing matrix is unique to its corresponding source, enabling data generation in a natural manner.

This study developed a new model, IC-U-Net, by combining the strengths of ICA and U-Net in EEG artifact removal. This model, based on the U-Net architecture with a loss function ensemble, was trained with mixtures of independent EEG sources. This study is novel in three ways. First and foremost, the proposed IC-U-Net model is the first EEG artifact removal method that uses the U-Net architecture. Second, IC-U-Net has an ensemble of loss functions, which allows it to simultaneously capture multiple signal variations in EEG recordings. This ensemble design can overcome a major limitation of the frequency principle (Xu *et al.*, 2019), which is that a deep neural network often fits target functions to low frequencies first and later to high frequencies. Third, the proposed method is a nonlinear mapping scheme that learns the latent properties of independent brain sources and automatically removes multiple artifacts, such as eye blinks/movements, muscle activities, and line/channel noises, from EEG signals.

The IC-U-Net code is available at https://github.com/roseDwayane/AIEEG.

## 2. IC-U-Net

This study developed a novel technique, IC-U-Net (Fig. 2), which creates a mapping between noisy and clean signals to reconstruct signals and remove artifacts. The proposed method, which is based on the U-Net architecture, was trained with brain and non-brain sources through minimizing an ensemble of loss functions. The robustness and feasibility of the proposed IC-U-Net were evaluated through one simulation study (Section 3) and three real-world EEG experiments (Section 4).

*2.1 Network architecture*

The deep denoising autoencoder (DDAE) aims to establish a correspondence between noisy inputs, $\widetilde{\mathbf{X}}$, and reconstructed outputs, $\mathbf{Y}$, through encoding and decoding processes, enabling the latter to approximate the clean targets, $\mathbf{X}$. The objective function, $\mathcal{L}$, is a reconstruction loss function determining the disparity between inputs and outputs that the DDAE tries to minimize to achieve the best artifact removal performance, which can be expressed as

$$E[\mathcal{L}(\mathbf{Y},\mathbf{X})] = E\left[\mathcal{L}\left(\text{De}\left(\text{En}(\widetilde{\mathbf{X}})\right),\mathbf{X}\right)\right]. \tag{1}$$

First, the encoder network (En) embeds the signals, $\widetilde{\mathbf{X}}$, into the latent space, where noises and non-noises are assumed to be well distinguishable. Then, the decoder network (De) obtains the reconstructed signals, $\mathbf{Y} = \text{De}\left(\text{En}(\widetilde{\mathbf{X}})\right)$. $E$ is the expectation over the distribution of $\mathcal{L}(\mathbf{Y},\mathbf{X})$.

These encoding and decoding processes are implemented using the DDAE and a U-Net-like network architecture (Fig. 2A), which mainly comprises a one-dimensional (1D) convolution neural network. In the encoder network, multiple layers of CBR blocks (i.e., convolution, batch normalization, and ReLU activation function (Glorot *et al.*, 2011)) and downsampling blocks attempt to extract increasingly more abstract representations of the input data by using the max-pool operator. To extract contextual information, the number of convolution filters doubles after each downsampling block. During decoding, the upsampling block with 1D transposed convolution (i.e., deconvolution) works with the CBR blocks, and the number of convolution filters halves after each upsampling block. To better reconstruct signals and recover spatial information, skip connections activate at the end of each decoder layer to concatenate the resultant decoded features with the corresponding encoded features. Figure 2A details the model structure and the hyperparameters, including kernel sizes, numbers of convolution and deconvolution filters, and stride.

*2.2 Loss function ensemble*

To capture the highly fluctuating structure of EEG signals, this work adopted an ensemble approach of forming multiple loss functions that could collaboratively gauge the difference between reconstructed and clean signals. This ensemble is a simple linear combination of four terms covering the amplitude, velocity, acceleration, and frequency components of EEG signals:

$$\mathcal{L}_{\text{ens}} = \frac{1}{\sum_i^4 \alpha_i} (\alpha_1 \mathcal{L}_{\text{amp}} + \alpha_2 \mathcal{L}_{\text{vel}} + \alpha_3 \mathcal{L}_{\text{acc}} + \alpha_4 \mathcal{L}_{\text{freq}}), \qquad (2)$$

where $\alpha_1, \alpha_2, \alpha_3$, and $\alpha_4$ denote the weights used to regulate the contribution of each term. Given that the channel number is $c$ and the data length is $t$, the loss from the amplitude component is

$$\mathcal{L}_{\text{amp}} = \frac{1}{c \times t} \sum_{i=1}^{c} \sum_{j=1}^{t} \text{MSE}(\mathbf{Y}(i,j), \mathbf{X}(i,j)), \qquad (3)$$

where MSE indicates the operator of the mean square error (i.e., L2 loss) between two arguments. The second loss deduced from the velocity component ($\mathcal{L}_{\text{vel}}$) is the MSE between the first-order differential estimates of **Y** and **X**. The third loss deduced from the acceleration component ($\mathcal{L}_{\text{acc}}$) is the MSE between the second-order differential estimates of **Y** and **X**. The fourth loss is generated from the following frequency component ($\mathcal{L}_{\text{freq}}$):

$$\mathcal{L}_{\text{freq}} = \frac{1}{c \times l} \sum_{i=1}^{c} \sum_{\xi=1}^{l} \text{MSE}(\mathbf{F}_{\mathbf{Y}}(i,\xi), \mathbf{F}_{\mathbf{X}}(i,\xi)), \qquad (4)$$

where $\mathbf{F_Y}(i, \xi)$ and $\mathbf{F_X}(i, \xi)$ represent the spectral estimates of $\mathbf{Y}$ and $\mathbf{X}$, respectively, in the range of 1–50 Hz with $l$ frequency bins. Each spectral estimate is the power spectral density converted by fast Fourier transformation, followed by z-score normalization.

*2.3 Model training using brain and non-brain sources*

We hypothesize that EEG artifacts can be removed if the proposed model can learn latent representations of clean EEG through the DDAE. To implement this learning, the clean targets and noisy inputs of the proposed U-Net model were mixtures of brain sources (denoted as mixB) and mixtures of brain and non-brain sources (denoted as mixBnB), respectively, decomposed by ICA. Specifically, as shown in Fig. 2B, ICA was first applied to training data, and then ICLabel (Pion-Tonachini *et al.*, 2019), an automated IC classifier, was used to distinguish the Brain IC from the six non-brain ICs (Muscle, Eye, Heart, Line Noise, Channel Noise, and Other). The selected $c_0 \leq c$ Brain ICs were then backprojected to the sensor level using the inverse of the corresponding unmixing matrix to synthesize the mixB, $\mathbf{X}$, as follows:

$$\mathbf{X} = [\widehat{\mathbf{A}}_{\text{brain}}, \mathbf{0}_{c \times (c-c_0)}] \begin{bmatrix} \mathbf{S}_{\text{brain}} \\ \mathbf{S}_{\text{non-brain}} \end{bmatrix}. \quad (5)$$

The time series, $\mathbf{S}_{\text{brain}} \in \mathbb{R}^{c_0 \times t}$, were $c_0$ ICs with probabilities of being brain activities, estimated by ICLabel, were more than 80%. The time series, $\mathbf{S}_{\text{non-brain}} \in \mathbb{R}^{(c-c_0) \times t}$, denotes the $c - c_0$ ICs with probabilities of less than 80% being brain activities. The weight matrix corresponding to Brain ICs, $\widehat{\mathbf{A}}_{\text{brain}} \in \mathbb{R}^{c \times c_0}$, was preserved, but the one corresponding to non-brain ICs was set as a zero matrix, $\mathbf{0} \in \mathbb{R}^{c \times (c-c_0)}$, to maximally eliminate non-brain activities in $\mathbf{X}$.

Given that ICLabel had classified the categories of all ICs in the training dataset, the mixBnB, $\widetilde{\mathbf{X}}$, could be synthesized by combining $\mathbf{X}$ with at least one of the artifact ICs. Consider $c_1$ Eye ICs (i.e., blinks and eye movements) as an example:

$$\widetilde{\mathbf{X}} = [\widehat{\mathbf{A}}_{\text{brain}}, \widehat{\mathbf{A}}_{\text{eye}}, \mathbf{0}_{c \times (c-c_0-c_1)}] \begin{bmatrix} \mathbf{S}_{\text{brain}} \\ \mathbf{S}_{\text{eye}} \\ \vdots \end{bmatrix}, \quad (6)$$

where $\widehat{\mathbf{A}}_{\text{eye}} \in \mathbb{R}^{c \times c_1}$ and $\mathbf{S}_{\text{eye}} \in \mathbb{R}^{c_1 \times t}$ are the time series and corresponding weight matrix of Eye ICs, respectively. By arbitrarily combining mixB with recognizable artifacts, the range of mixBnB can be feasibly diversified to facilitate model training.

*2.4 Performance evaluation*

The proposed model was validated using the MSE and SNR as the performance metrics (Eqs. (3) and (4)) in one simulation (Section 3) and three real-world EEG experiments (Section 4). Specifically, the MSE between the mixB, $\mathbf{X}$, and the reconstructed output, $\mathbf{Y}$, was used to evaluate the model performance and confirm the convergence of the model training, and the SNR was calculated as

$$\text{SNR} = \frac{1}{c}\sum_{i=1}^{c} 10 \times \log\left(\frac{\sum_{j=1}^{t}(\mathbf{X}(i,j))^2}{\sum_{j=1}^{t}(\mathbf{Y}(i,j) - \mathbf{X}(i,j))^2}\right). \quad (7)$$

These two metrics were further averaged across samples, with a lower MSE and a higher SNR indicating better artifact removal and signal reconstruction performance.

Two other validation methods were used to determine whether the reconstructed signals were brain activities. First, the number of brain sources was calculated. Specifically, ICA and ICLabel were applied again, but this time to the reconstructed signals, to calculate the numbers of available Brain

ICs. We hypothesize that an effective artifact removal method can increase the number of brain sources in the reconstructed signals. Second, within the reconstructed EEG signals, the presence of typical EEG activity, such as event-related potential (ERP), in response to specific events was verified.

## 3. Simulation Experiments and Results

The proposed method was first validated through a simulation experiment (Fig. 3A) to verify 1) the convergence of model training, 2) the preference of each loss function in capturing frequency bins, and 3) the robustness of signal reconstruction.

*3.1 Simulated signals and model hyperparameters*

This experiment used a synthetic dataset of 10,240 samples (8,192 (80%) training samples, 1,024 (10%) test samples, and 1,024 (10%) validation samples). The dimensions of each synthetic time series were 19 × 1,024, representing 4-s of 19-channel data sampled at 256 Hz. Each time series was made of a linear combination of six sinusoidal waves whose frequency (Hz), amplitude (arbitrary units), and phase (radians) ranged between (0,50), (0,1), and (0,2π), respectively. Z-score normalization was used to rescale each synthetic time series. The model hyperparameters, maximum number of epochs, batch size, and learning rate were set to 150, 128, and 0.01, respectively.

*3.2 Ablation study of loss functions*

Loss functions can have a big impact on a model's performance. Thus, this section examines the convergence of model training and the capability of each loss function in capturing frequency. Figure 4A compares the loss values of the validation samples derived from the model trained using (left to right) loss functions $\mathcal{L}_{amp}$ ($\boldsymbol{\alpha} = [\alpha_1, \alpha_2, \alpha_3, \alpha_4] = [1,0,0,0]$), $\mathcal{L}_{vel}$ ($\boldsymbol{\alpha} = [0,1,0,0]$), $\mathcal{L}_{acc}$ ($\boldsymbol{\alpha} = [0,0,1,0]$), $\mathcal{L}_{freq}$ ($\boldsymbol{\alpha} = [0,0,0,1]$), and $\mathcal{L}_{ens}$ ($\boldsymbol{\alpha} = [1,1,1,1]$). As the number of epochs increased, all loss values (bold traces) decreased, and all SNRs increased (Fig. 4B), indicating that the models trained with different loss functions could all reach convergence. All loss values of the models trained with $\mathcal{L}_{amp}$ (bold green trace), $\mathcal{L}_{vel}$ (bold orange trace), $\mathcal{L}_{acc}$ (bold yellow trace), and $\mathcal{L}_{ens}$ (bold magenta trace) decreased to less than 0.01, with the only exception being the loss values of the model trained with $\mathcal{L}_{freq}$ (bold purple trace). Overall, the model trained with the ensemble of loss functions, $\mathcal{L}_{ens}$, outperformed the others, achieving the lowest loss value (<0.003), the most stable loss value change (<$10^{-4}$), and the highest SNR (>25 dB).

These loss values provide insights into the couplings between various loss functions. Take $\boldsymbol{\alpha} = [1,0,0,0]$ as an example, the three non-bold traces in Fig. 4A-i are the changes of $\mathcal{L}_{vel}$, $\mathcal{L}_{acc}$, and $\mathcal{L}_{freq}$ that were deactivated in the model trained with only $\mathcal{L}_{amp}$. These loss values followed the same decreasing trend as the bold green trace, indicating that minimizing amplitude errors would also facilitate the capturing of the velocity, acceleration, and frequency components from the signals. Additionally, minimizing $\mathcal{L}_{vel}$ and $\mathcal{L}_{acc}$ would also reduce the errors in the acceleration (Fig. 4A-ii) and velocity (Fig. 4A-iii) components, respectively. Minimizing $\mathcal{L}_{freq}$ did capture the frequency component (Fig. 4A-iv) but failed to minimize the errors in the amplitude, velocity, and acceleration components. Regarding the loss function ensemble, the proposed model trained with $\mathcal{L}_{\text{ens}}$ (Fig. 4A-v) could jointly decrease all terms in Eq. (2) within comparable numbers of epochs to reach convergence.

*3.3 Simulated signal reconstruction*

Figure 4C presents an example of a 1-s temporal segment reconstructed using the proposed model with different loss functions. The models trained with $\mathcal{L}_{\text{amp}}$ and $\mathcal{L}_{\text{ens}}$ had the greatest results in terms of restoring the waves and the frequency components and the smallest errors across frequency bins (Fig. 4D). The error distribution suggested that $\mathcal{L}_{\text{vel}}$ and $\mathcal{L}_{\text{acc}}$ could preferentially capture high-frequency components over low-frequency components. Although the errors obtained from the models trained with $\mathcal{L}_{\text{vel}}$, $\mathcal{L}_{\text{acc}}$, and $\mathcal{L}_{\text{freq}}$ were larger than those obtained from the model trained with $\mathcal{L}_{\text{amp}}$, by combining all loss functions, the model was still able to suppress the overall errors across frequency bins. Further investigation of the error peak between 30 and 35 Hz is required.

Table 1 summarizes the best MSE and SNR obtained using various loss functions. The model trained with $\mathcal{L}_{\text{amp}}$ outperformed the models trained with the other three loss functions. The best signal reconstruction was achieved by the model trained with the ensemble of loss functions, as evidenced by this model having the lowest MSE and highest SNR.

## 4. Real-world EEG Experiments and Results

The proposed IC-U-Net model was trained using a large number of real resting-state EEG signals from 546 recording sessions. As in the simulations, loss values and SNR were used to characterize the model's convergence and performance. Moreover, the model's performance was assessed against a variety of artifacts (Fig. 3B). The proposed model was further validated against two non-resting-state EEG datasets collected in a lane-keeping driving experiment (Cao *et al.*, 2019) (Section 4.3) and a walking experiment (Section 4.4). To verify that the reconstructed signals were brain activities and still possessed the characteristics of EEG signals, after completing artifact removal processing, ICA and ICLabel (Pion-Tonachini *et al.*, 2019) were applied to the reconstructed signals to see if the number of non-brain ICs decreased and whether the number of extractable brain ICs increased.

The artifact removal performance of the proposed model was compared to that of three other methods (Fig. 3C): bandpass filter (frequency range = 1–50 Hz), a component-based method (i.e., ASR (Chang *et al.*, 2020) with burst cutoff parameter $k = 5$), and a DL-based method (i.e., 1D-ResCNN (Sun *et al.*, 2020) with EEGdenoiseNet (Zhang *et al.*, 2021) and hyperparameters set to default values).

*4.1 IC-U-Net model building*

A database of 546 EEG datasets [specifically 438 (~80%) training datasets, 54 (~10%) validation datasets, and 54 (~10%) test datasets] was used to build the proposed IC-U-Net model. Each dataset was collected from a human subject during a nearly 5-min resting-state 30-channel EEG (Nuamps,

Compumedics Neuroscan) acquisition session with their eyes open and closed. The electrodes were placed according to the extended international 10–20 system at FP1, FP2, F7, F3, Fz, F4, F8, FT7, FC3, FCz, FC4, FT8, T3, C3, Cz, C4, T4, TP7, CP3, CPz, CP4, TP8, T5, P3, Pz, P4, T6, O1, Oz, and O2. The sampling rate was set to 1,000 Hz, which was then downsampled to 256 Hz.

After being decomposed into brain and artifact sources by ICA and ICLabel, each dataset was segmented into multiple 4-s segments [each of dimension 30 (channels) × 1,024 (points)] in a non-overlapping manner. Following the procedures described in Section 2.3, this study generated clean targets (i.e., mixB) by mixing brain sources through the corresponding weights of spatial distribution and noisy data (i.e., mixBnB) by mixing brain sources with one of the artifacts (Eye, Muscle, Heart, Channel Noise, and Other). Table 2 lists the numbers of mixB and mixBnB records; Line Noise was not present in any of the datasets. Before training, z-score normalization was used to rescale each time series. As the model hyperparameters, the number of epochs, batch size, and learning rate were set to 150, 128, and 0.01, respectively.

*4.2 Model validation*

As shown in Figs. 5A and 5B, the validation loss decreased and the SNR increased as the epoch size increased, thereby supporting the feasibility of the proposed model in processing real-world EEG data. Similar to the simulation results in Section 3, the model trained with the ensemble of loss functions (i.e., $\alpha = [1,1,1,1]$) achieved the best performance. Figure 5C shows four 4-s EEG time series heavily contaminated by Eye, Muscle, Channel Noise, and Other (gray traces) noises and the corresponding time series reconstructed by IC-U-Net. The reconstructed signals (magenta traces) resembled mixB (black traces) in the time and frequency domains. The low-frequency components caused by Eye and Other noises were removed, whereas the high-frequency components (Muscle noise) were alleviated; the broadband frequency components occurring in signals with Channel Noise were also suppressed.

Table 3 lists the MSE and SNR obtained using the proposed model trained with different loss functions. The model trained with $\mathcal{L}_{\text{amp}}$ outperformed the models trained with the other three loss functions. As expected, the model trained with the ensemble of loss functions, $\mathcal{L}_{\text{ens}}$, exhibited the best performance.

*4.3 Model deployment I: lane-keeping driving experiment*

The proposed model was then tested on one of our existing databases collected in a virtual driving experiment (Chuang *et al.*, 2014; Lin *et al.*, 2016) to see if it can remove pervasive artifacts during driving and compare its performance to that of other methods (Fig. 3C). In the experiment, each subject participated in a 1.5-h event-related lane departure task in the afternoon, during which time their 30-channel EEG signals were collected at a sampling rate of 500 Hz using Neuroscan SynAmps2. The electrode placement was identical to that in the resting-state experiment described in Section 4.1. The objective of this experiment was to investigate the EEG correlates of sustained attention during driving. The subjects were tasked with keeping the vehicle cruising in the center of the designated lane by correcting the vehicle's trajectory (using the steering wheel) as soon as possible if the vehicle began to drift. Seventy-six test datasets were selected in total, 60 of which are publicly available (Cao *et al.*, 2019) and 16 were added from our data repository. Before testing, the data were downsampled to 256 Hz and filtered using a zero-phase FIR filter with a cutoff frequency between 0.1 and 80 Hz.

Figures 6A-6C showed that ASR, 1D-ResCNN, and IC-U-Net could all remove artifacts from EEG signals with a low MSE and a high SNR (Table 4). However, only IC-U-Net could best preserve the waveform for a clean EEG segment. Overall, the proposed IC-U-Net model outperformed the other three. To further investigate whether the reconstructed signals were brain activities, this study analyzed ERPs time-locked to lane-departure events to verify the presence of neural responses associated with attention processing, with the assumption that the classical P3 activity would occur in response to the stimulus-driven event (Bledowski *et al.*, 2004). As revealed by the ERP data (Delorme and Makeig, 2004) (upper panel) and their average (lower panel) in Fig. 6D, the signals processed with the filter, ASR, and IC-U-Net models preserved clear P3 amplitudes.

Next, this study analyzed the components underlying the reconstructed signals to confirm that applying an artifact removal method to EEG can facilitate the extraction and preservation of relatively more brain activities. As shown in Fig. 6E, the proposed method and the three representative methods could all extract more Brain sources from the reconstructed signals (color circles) than from the raw signals (gray circles). Specifically, IC-U-Net allowed the ICA to extract the most (i.e., 23.9 ± 2.8) Brain sources from 30-channel EEG signals. In addition, the extracted Brain sources contained an average of >80% activity arising from cortical patches (Table 5). Furthermore, Muscle, Line Noise, and Channel Noise were all removed from the signals processed with the proposed model. In summary, IC-U-Net outperformed the other three models.

*4.4 Model deployment II: walking experiment*

Finally, the proposed model was tested in a walking experiment (Fig. 7A). An adult subject's 32-channel EEG activities (LiveAmp, Brain Products GmbH) were recorded while walking down a jogging track in a stadium. Figure 7B shows the EEG time series before and after artifact removal using IC-U-Net. ICA and ICLabel were used to analyze the components in the reconstructed signal. As shown in Fig. 7C, IC-U-Net removed 16 artifact sources—4 Eye, 1 Channel Noise, and 11 Line Noise—from the original EEG recording. There was only one identified Brain source (>80% brain) in the original EEG recording, whereas after reconstruction using IC-U-Net, the number of qualified Brain sources increased substantially to 16. Table 6 summarizes the probability of finding decomposed brain sources in the reconstructed signals; these results suggest that IC-U-Net could increase the probability of finding brain sources in walking data from 4.5% to 61.8%. Such a remarkable improvement was also found in resting-state and lane-keeping driving data. Taken together, the results validate the use of the proposed IC-U-Net method for reconstructing EEG signals to increase the number of extractable brain sources.

## 5. Conclusion

This study proposed a new end-to-end artifact removal method, IC-U-Net, based on the DDAE architecture of U-Net, the source separation technique of ICA, the IC classification of ICLabel, and an ensemble of loss functions. The proposed IC-U-Net model, which we have made publicly available, was trained with over 150,000 segments of 4-s mixtures of resting-state EEG ICs. The results of one simulation and three experiments based on real-world EEG data demonstrated that IC-U-Net can automatically remove various artifacts from EEG recordings and effectively increase the number of extractable brain sources. The user-friendly IC-U-Net has three distinct advantages. First, IC-U-Net takes raw EEG time series as inputs and outputs artifact-corrected time series. The entire signal reconstruction procedure is done at the sensor level, without the need to tune any parameters. The most time-consuming step, ICA, is only involved in the data synthesis and model validation. Second, users do not need to specify the types of noises to be removed. IC-U-Net recovers mixtures of brain sources while suppressing non-brain sources, allowing it to remove multiple artifacts at once. Third, IC-U-Net can process EEG data with any number of channels. Overall, these results demonstrated the superiority of IC-U-Net as a preprocessing tool for EEG analysis and BCI applications.

**Table 1.** Performance of the proposed IC-U-Net model on the simulation dataset

| Data | Loss functions | MSE | | | SNR | | |
|---|---|---|---|---|---|---|---|
| | | Training | Validation | Test | Training | Validation | Test |
| Simulation experiment | $\mathcal{L}_{amp}$ (epoch*=109) | 0.03 ±0.02 | 0.03 ±0.02 | 0.03 ±0.04 | 22.60 ±2.48 | 22.52 ±2.58 | 22.50 ±2.55 |
| | $\mathcal{L}_{vel}$ (epoch*=110) | 0.47 ±0.30 | 0.49 ±0.30 | 0.49 ±0.33 | 7.76 ±0.79 | 7.74 ±0.82 | 7.73 ±0.82 |
| | $\mathcal{L}_{acc}$ (epoch*=112) | 0.66 ±0.67 | 0.67 ±0.65 | 0.70 ±0.77 | 7.60 ±3.44 | 7.67 ±3.45 | 7.54 ±3.49 |
| | $\mathcal{L}_{freq}$ (epoch*=141) | 4.57 ±2.68 | 4.66 ±2.68 | 4.61 ±2.85 | -1.14 ±1.18 | -1.18 ±1.10 | -1.21 ±1.13 |
| | $\mathcal{L}_{ens}$ (epoch*=119) | **0.01** ±0.01 | **0.02** ±0.01 | **0.02** ±0.02 | **24.98** ±2.45 | **24.88** ±2.59 | **24.88** ±2.58 |

* The epoch number at which the best result is achieved.

**Table 2.** Number of 4-s resting-state EEG segments used to build IC-U-Net

| Category | Mixture of Sources | Training | Validation | Test |
|---|---|---|---|---|
| mixB | Brain | 39,605 | 3,592 | 3,693 |
| mixBnB | Brain + Eye | 33,369 | 2,982 | 3,337 |
| | Brain + Muscle | 27,429 | 2,416 | 2,534 |
| | Brain + Heart | 1,498 | 75 | 151 |
| | Brain + Channel Noise | 12,686 | 1,380 | 1,415 |
| | Brain + Line Noise | N/A | N/A | N/A |
| | Brain + Other | 36,234 | 3,515 | 3,636 |
| | Total | 150,821 | 13,960 | 14,766 |

N/A indicates that none of the Line Noise is extractable from the resting-state EEG datasets.

Refer to ICLabel for the definitions of Eye, Muscle, Heart, Channel Noise, Line Noise, and Other.

**Table 3.** Performance of the proposed IC-U-Net model on the resting-state datasets

| Dataset | Loss functions | MSE | | | SNR | | |
|---|---|---|---|---|---|---|---|
| | | Training | Validation | Test | Training | Validation | Test |
| Resting-state with eyes open and closed | $\mathcal{L}_{\text{amp}}$ (epoch*=118) | 0.39 ±0.26 | 0.40 ±0.21 | 0.42 ±0.27 | 5.61 ±2.51 | 5.09 ±1.97 | 5.28 ±2.61 |
| | $\mathcal{L}_{\text{vel}}$ (epoch*=104) | 1.40 ±0.23 | 1.43 ±0.25 | 1.43 ±0.24 | -1.48 ±0.86 | -1.66 ±1.04 | -1.53 ±0.89 |
| | $\mathcal{L}_{\text{acc}}$ (epoch*=108) | 1.70 ±0.09 | 1.70 ±0.09 | 1.71 ±0.10 | -3.46 ±0.44 | -3.54 ±0.32 | -3.46 ±0.51 |
| | $\mathcal{L}_{\text{freq}}$ (epoch*=126) | 1.83 ±0.13 | 1.82 ±0.09 | 1.83 ±0.12 | -2.29 ±0.51 | -2.55 ±0.30 | -2.29 ±0.54 |
| | $\mathcal{L}_{\text{ens}}$ (epoch*=145) | **0.37** ±0.27 | **0.38** ±0.20 | **0.40** ±0.29 | **6.03** ±2.66 | **5.49** ±2.08 | **5.68** ±2.77 |

* The epoch number at which the best result is achieved.

**Table 4.** Performance comparison of the proposed method and three other methods

| Dataset | | Artifact* | | | | | |
|---|---|---|---|---|---|---|---|
| | | Eye | Muscle | Other | Eye | Muscle | Other |
| | Number | 171 | 273 | 273 | 171 | 273 | 273 |
| | Methods | MSE | | | SNR | | |
| Lane-keeping driving datasets | Filter | 1.81 ±1.08 | 1.74 ±0.74 | 1.63 ±0.78 | -2.25 ±2.97 | -2.52 ±3.37 | -2.14 ±3.32 |
| | ASR | 1.44 ±0.98 | 1.40 ±0.89 | 1.44 ±0.99 | -1.06 ±3.51 | -1.29 ±3.58 | -1.36 ±3.12 |
| | 1D-ResCNN | 1.42 ±0.94 | 2.12 ±0.93 | 1.42 ±0.73 | -0.98 ±2.66 | -3.31 ±3.24 | -1.43 ±2.64 |
| | IC-U-Net | **0.57** ±0.58 | **0.56** ±0.51 | **0.87** ±0.55 | **4.06** ±3.56 | **3.62** ±3.09 | **1.16** ±3.34 |

*Heart, Line Noise, and Channel Noise are omitted because they are rarely decomposed from this dataset.

**Table 5.** Number of sources decomposed from reconstructed EEG signals

| | Decomposed Sources | Quantity | Probability (in %) | | | | | | |
|---|---|---|---|---|---|---|---|---|---|
| | | | Brain | Muscle | Eye | Heart | Line Noise | Channel Noise | Other |
| Raw | | 7.8 ±3.2 | 71.4 ±9.8% | 1.6 ±2.0% | 1.2 ±1.3% | 0.7 ±1.0% | 7.4 ±6.9% | 2.2 ±1.7% | 15.5 ±5.8% |
| Filter | | 13.1 ±3.9 | 81.3 ±6.8% | 2.6 ±2.1% | 0.9 ±1.3% | 0.5 ±0.5% | 1.1 ±0.7% | 1.7 ±1.2% | 11.7 ±4.8% |
| ASR | Brain | 17.8 ±3.0 | 87.4 ±4.6% | 1.7 ±1.3% | 0.7 ±0.7% | 0.5 ±0.6% | 0.7 ±0.4% | 1.2 ±0.9% | 7.9 ±3.1% |
| 1D-ResCNN | | 19.7 ±3.0 | 83.2 ±5.2% | 2.0 ±1.1% | 0.5 ±0.4% | 0.3 ±0.2% | 1.5 ±1.4% | 1.7 ±0.9% | 10.8 ±3.4% |
| IC-U-Net | | **23.9 ±2.8** | 83.9 ±4.9% | 0.5 ±0.3% | 0.6 ±0.5% | 1.2 ±0.8% | 0.6 ±0.3% | 0.7 ±0.4% | 12.5 ±4.2% |
| Raw | | 2.4 ±1.5 | 70.8 ±0.5% | 65.7 ±14.0% | 2.6 ±2.7% | 0.7 ±0.8% | 2.2 ±5.3% | 3.2 ±3.3% | 18.6 ±8.9% |
| Filter | | 4.1 ±2.6 | 8.3 ±5.6% | 68.0 ±12.5% | 2.3 ±3.3% | 0.7 ±0.6% | 0.5 ±0.4% | 4.2 ±6.0% | 16.0 ±8.1% |
| ASR | Muscle | 4.2 ±2.5 | 7.3 ±4.6% | 70.2 ±11.7% | 2.8 ±3.7% | 0.6 ±0.5% | 0.5 ±0.3% | 4.0 ±3.2% | 14.6 ±8.4% |
| 1D-ResCNN | | 1.3 ±0.5 | 17.3 ±8.7% | 41.5 ±10.0% | 3.2 ±5.3% | 0.9 ±0.7% | 1.0 ±1.3% | 19.5 ±9.6% | 16.6 ±6.9% |
| IC-U-Net | | − | − | − | − | − | − | − | − |
| Raw | | 4.4 ±1.8 | 4.1 ±3.4% | 1.5 ±2.6% | 79.8 ±8.6% | 0.5 ±0.4% | 1.9 ±2.7% | 2.9 ±2.1% | 9.4 ±4.9% |
| Filter | | 3.8 ±1.5 | 3.8 ±3.1% | 3.5 ±3.6% | 77.6 ±9.1% | 0.6 ±0.7% | 0.2 ±0.2% | 2.7 ±2.5% | 11.5 ±5.5% |
| ASR | Eye | 2.4 ±0.8 | 6.0 ±5.4% | 2.7 ±5.5% | 84.0 ±10.2% | 0.2 ±0.2% | 0.1 ±0.1% | 0.8 ±1.3% | 6.1 ±4.9% |
| 1D-ResCNN | | 1.6 ±0.7 | 5.2 ±4.5% | 4.0 ±4.1% | 59.4 ±15.4% | 0.3 ±0.4% | 1.2 ±4.5% | 8.3 ±6.3% | 21.6 ±8.6% |
| IC-U-Net | | 1.0 ±0.0 | 21.1 ±13.1% | 0.4 ±0.4% | 56.0 ±14.4% | 2.9 ±3.6% | 0.2 ±0.2% | 1.9 ±4.3% | 17.5 ±13.0% |
| Raw | | 1.0 ±0.0 | 14.6 ±6.8% | 1.1 ±0.8% | 1.2 ±0.1% | 45.8 ±23.4% | 10.7 ±13.2% | 1.3 ±0.4% | 25.1 ±4.7% |
| Filter | | 1.7 ±1.2 | 15.0 ±12.9% | 2.2 ±1.0% | 0.3 ±0.3% | 60.8 ±2.4% | 0.6 ±0.3% | 1.5 ±0.4% | 19.6 ±14.4% |
| ASR | Heart | 1.0 ±0.0 | 12.8 ±13.6% | 9.6 ±0.2% | 0.5 ±0.2% | 40.9 ±5.8% | 2.3 ±2.6% | 1.4 ±1.2% | 32.5 ±3.6% |
| 1D-ResCNN | | − | − | − | − | − | − | − | − |
| IC-U-Net | | 1.0 ±0.0 | 22.8 ±9.6% | 0.8 ±0.5% | 0.1 ±0.1% | 66.0 ±11.2% | 0.3 ±0.2% | 1.6 ±1.3% | 8.3 ±6.2% |
| Raw | | 5.1 ±3.9 | 13.0 ±7.0% | 0.7 ±1.6% | 2.2 ±3.0% | 0.5 ±0.8% | 60.4 ±10.4% | 3.1 ±2.7% | 20.1 ±9.3% |
| Filter | | 1.0 ±0.0 | 10.9 ±0.0% | 2.8 ±0.0% | 0.2 ±0.0% | 0.7 ±0.0% | 36.6 ±0.0% | 23.2 ±0.0% | 25.5 ±0.0% |
| ASR | Line Noise | 1.0 ±0.0 | 6.9 ±0.0% | 0.0 ±0.0% | 0.0 ±0.0% | 0.0 ±0.0% | 56.7 ±0.0% | 0.2 ±0.0% | 36.2 ±0.0% |
| 1D-ResCNN | | 1.7 ±1.0 | 16.4 ±9.6% | 0.1 ±0.2% | 0.5 ±0.8% | 0.2 ±0.3% | 69.1 ±13.0% | 0.7 ±0.9% | 12.9 ±11.4% |
| IC-U-Net | | − | − | − | − | − | − | − | − |
| Raw | | 1.4 ±0.6 | 15.0 ±7.3% | 5.7 ±7.7% | 4.0 ±5.3% | 0.8 ±0.5% | 7.3 ±9.4% | 51.7 ±14.6% | 15.6 ±8.4% |
| Filter | | 1.2 ±0.6 | 15.9 ±8.9% | 8.3 ±6.7% | 1.9 ±5.8% | 0.7 ±0.5% | 1.1 ±0.7% | 52.1 ±14.8% | 20.0 ±9.0% |
| ASR | Channel Noise | 1.3 ±0.5 | 20.1 ±9.2% | 8.6 ±9.0% | 1.3 ±0.8% | 0.8 ±0.5% | 2.0 ±4.1% | 49.1 ±11.7% | 18.1 ±8.7% |
| 1D-ResCNN | | 1.2 ±0.6 | 15.3 ±7.7% | 21.5 ±11.9% | 2.7 ±4.1% | 1.2 ±0.9% | 1.1 ±1.1% | 42.0 ±8.9% | 16.2 ±9.9% |
| IC-U-Net | | − | − | − | − | − | − | − | − |
| Raw | | 13.5 ±4.6 | 14.6 ±4.3% | 3.2 ±2.3% | 3.4 ±2.4% | 1.1 ±0.8% | 7.5 ±5.8% | 4.4 ±2.2% | 65.6 ±7.2% |
| Filter | | 9.2 ±4.2 | 16.0 ±5.4% | 8.1 ±3.9% | 3.1 ±2.6% | 1.2 ±1.0% | 1.9 ±0.8% | 4.9 ±2.7% | 64.9 ±7.7% |
| ASR | Other | 4.9 ±2.5 | 17.4 ±7.2% | 9.6 ±6.7% | 1.9 ±2.8% | 1.2 ±1.3% | 1.9 ±1.6% | 4.0 ±3.4% | 64.1 ±7.2% |
| 1D-ResCNN | | 7.7 ±3.3 | 20.3 ±5.0% | 5.4 ±2.4% | 2.9 ±2.3% | 0.7 ±0.5% | 3.1 ±2.0% | 5.9 ±3.0% | 61.8 ±7.4% |
| IC-U-Net | | 6.0 ±2.8 | 25.9 ±6.9% | 1.1 ±0.8% | 1.4 ±1.6% | 3.1 ±2.5% | 1.2 ±0.7% | 1.9 ±1.5% | 65.3 ±8.3% |

− indicates that none of the sources could be decomposed after signal reconstruction.

**Table 6.** Average probability of brain activity across all decomposed ICs in reconstructed EEG signals

|  |  | Resting-state | | Lane-keeping driving | Walking |
|---|---|---|---|---|---|
|  |  | Training + Validation data | Test data | | |
| EEG device | | NuAmp | | SynAmps2 | LiveAmp |
| Channel number | | 30 | | 30 | 30 |
| Sampling rate (Hz) | | 500→256 | | 500→256 | 1000→256 |
| Data size (session) | | 438+54 | 54 | 76 | 1 |
| Data length (sec) | | 289.3±32.9 | 290.0±32.4 | 4,737.3±1,266.9 | 1,016.00 |
| Probability of Brain in all ICs (%) | Raw | 53.9±12.7 | 53.8±10.4 | 27.6±8.6 | 4.5 |
|  | Filter | 56.8±12.0 | 57.1±9.6 | 42.2±10.5 | 13.2 |
|  | ASR | 64.5±10.0 | 64.1±8.5 | 56.4±8.8 | 16.9 |
|  | 1D-ResCNN | 69.7±9.2 | 69.1±11.4 | 60.6±8.2 | 36.1 |
|  | IC-U-Net | **90.3±9.2** | **89.1±6.9** | **72.1±8.8** | **61.8** |

→ indicates downsampling.

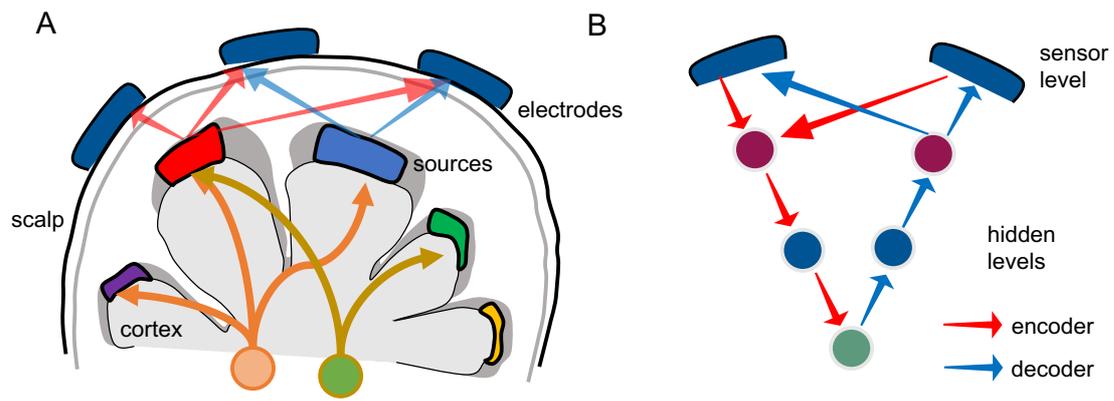

Figure 1. Illustration of (A) source separating and mixing processes, modified from (Onton and Makeig, 2009), and (B) architecture of the encoder and decoder.

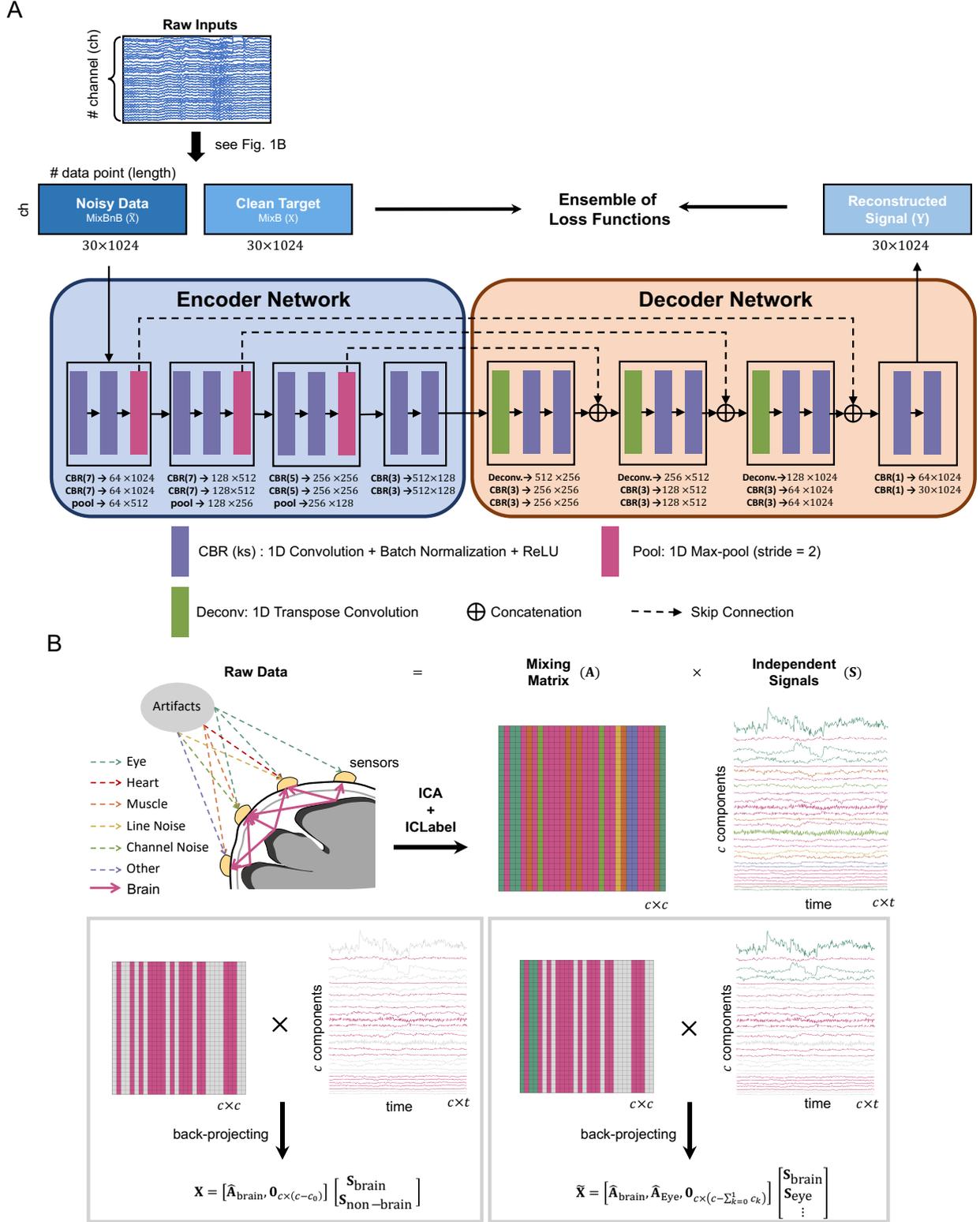

Figure 2. IC-U-Net. The proposed model was built based on (A) the U-Net architecture with an ensemble of loss functions and trained with (B) mixtures of independent sources (i.e., mixB and mixBnB) at the sensor level.

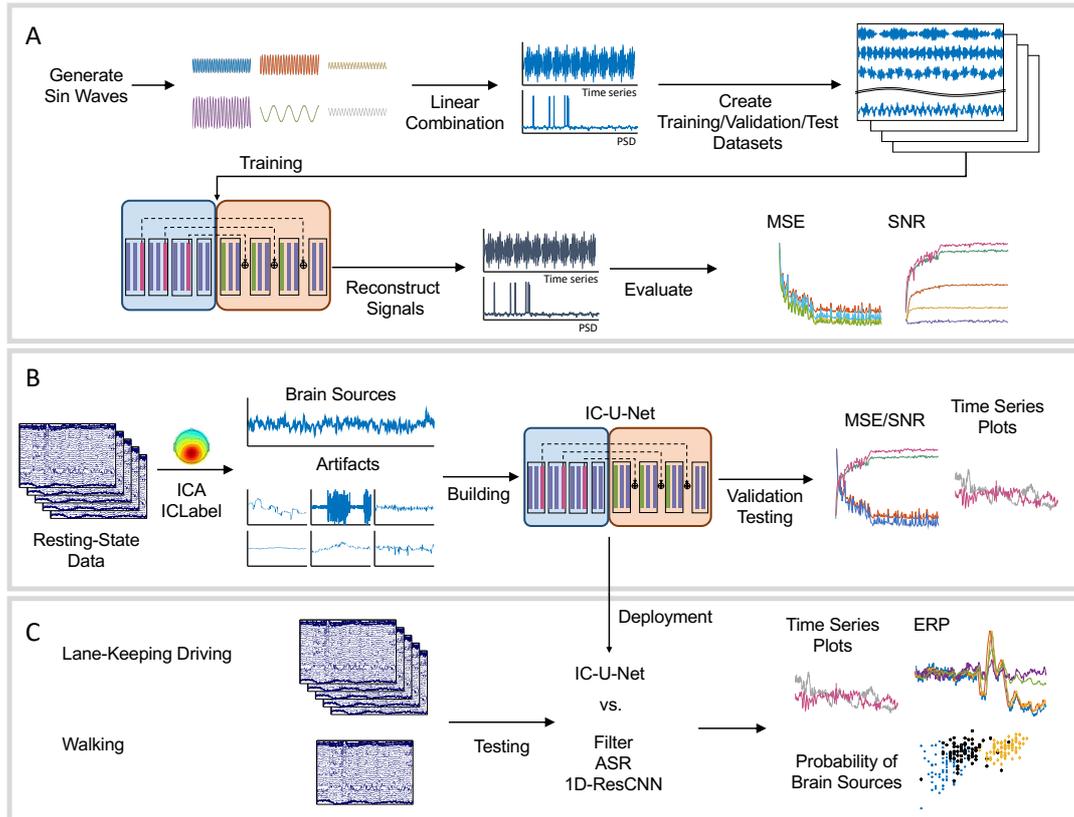

Figure 3. Model training and validation. Performance was validated through (A) a simulation experiment and three real-world EEG datasets collected during (B) resting state and (C) lane-keeping and walking experiments. Various metrics and assessments (i.e., the MSE, SNR, time series plot, ERP, and probability of brain sources) are calculated to evaluate the convergence of the model training, the effectiveness of different loss functions in capturing signal fluctuations, the effectiveness of artifact removal, and the effectiveness of brain source reconstruction.

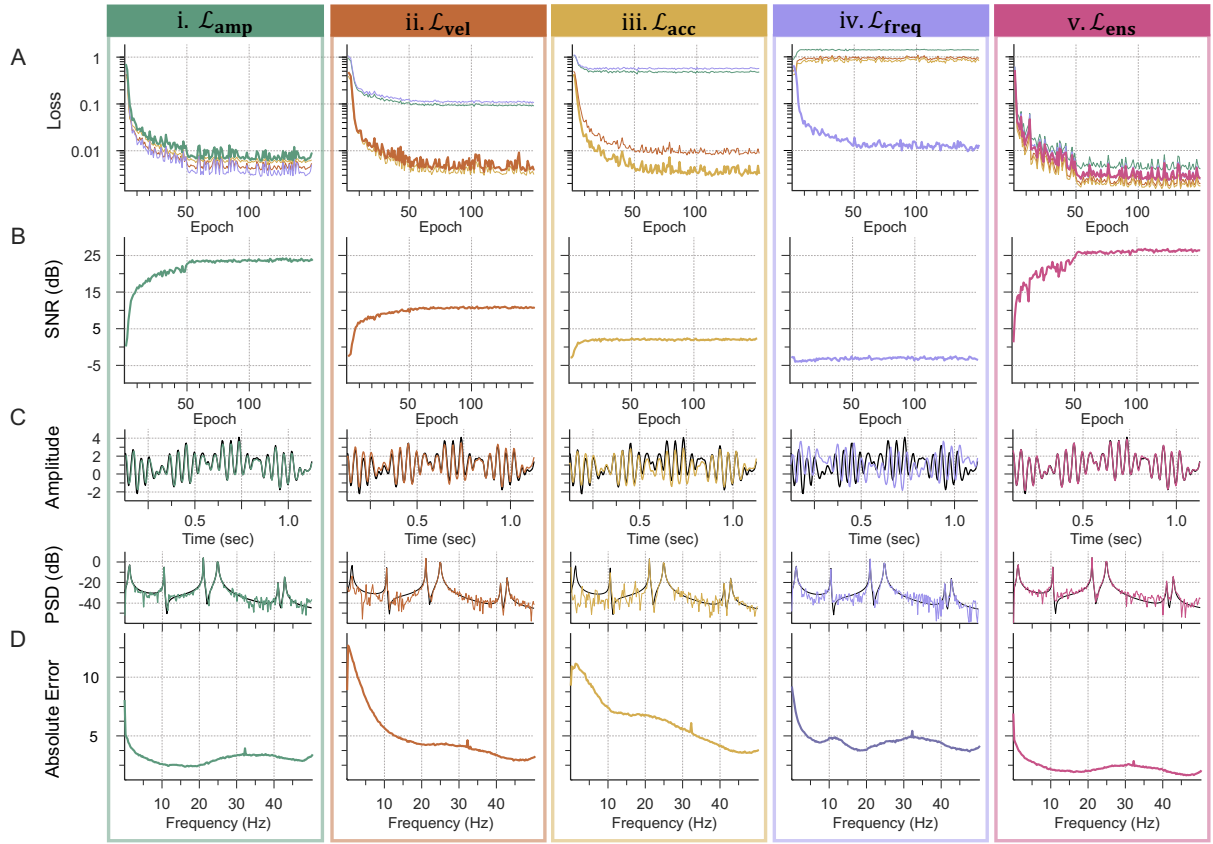

Figure 4. Ablation study of loss functions in the simulation experiment. (A and B) Changes in validation loss and SNR as epoch size increases. (C) A time series of power spectral density (PSD) of the simulated data before (black traces) and after (color traces) signal recovery. The simulated data are signals of 1.05, 10.44, 20.97, 24.88, 41.05, and 42.81 Hz. (D) The average absolute errors of each frequency bin over all test samples.

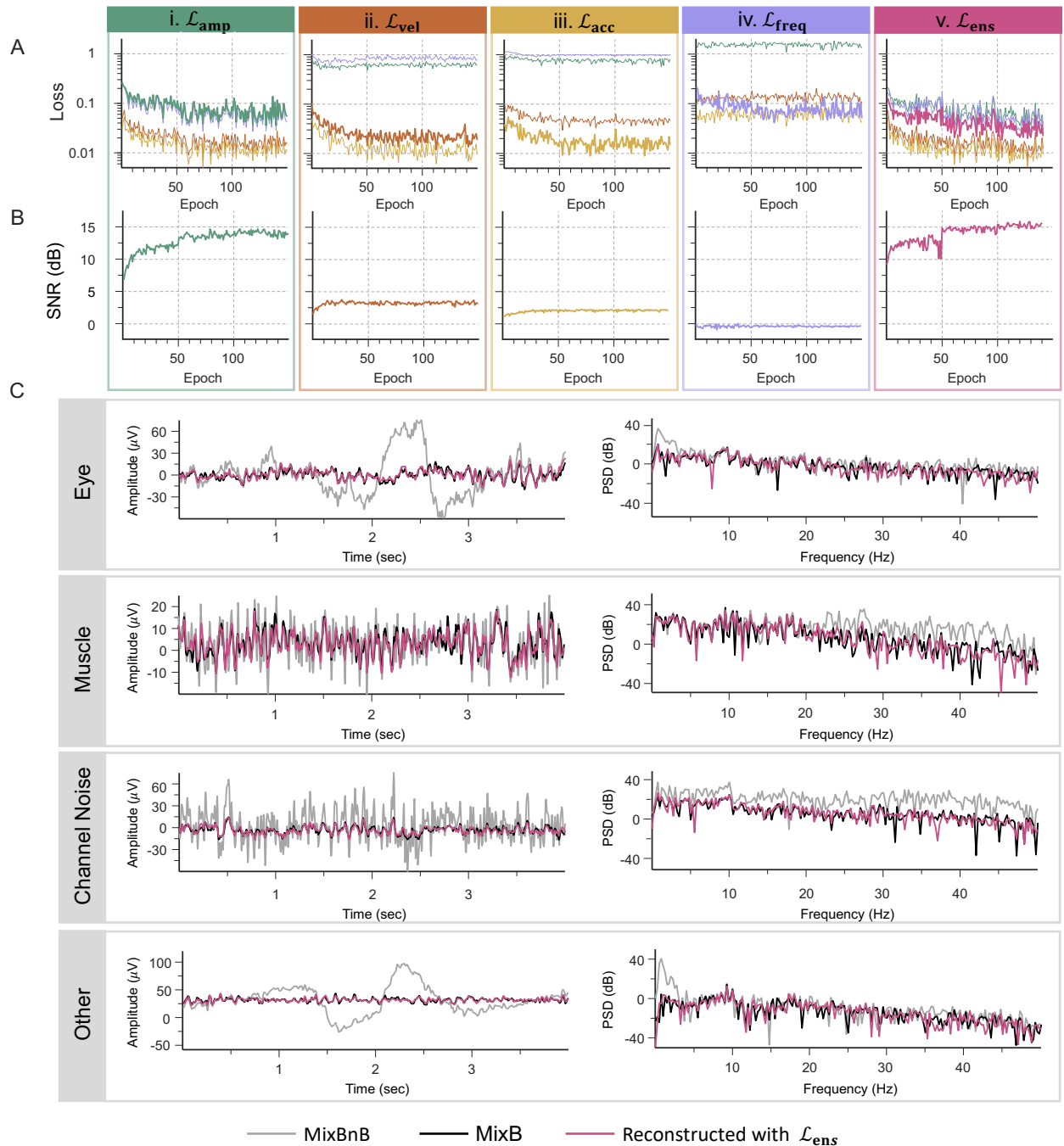

Figure 5. IC-U-Net model training. (A and B) Changes in validation loss and SNR with increase in epoch size. (C) Recovery of 4-s segments of MixB contaminated by Eye, Muscle, Channel Noise, and Other ICs using IC-U-Net.

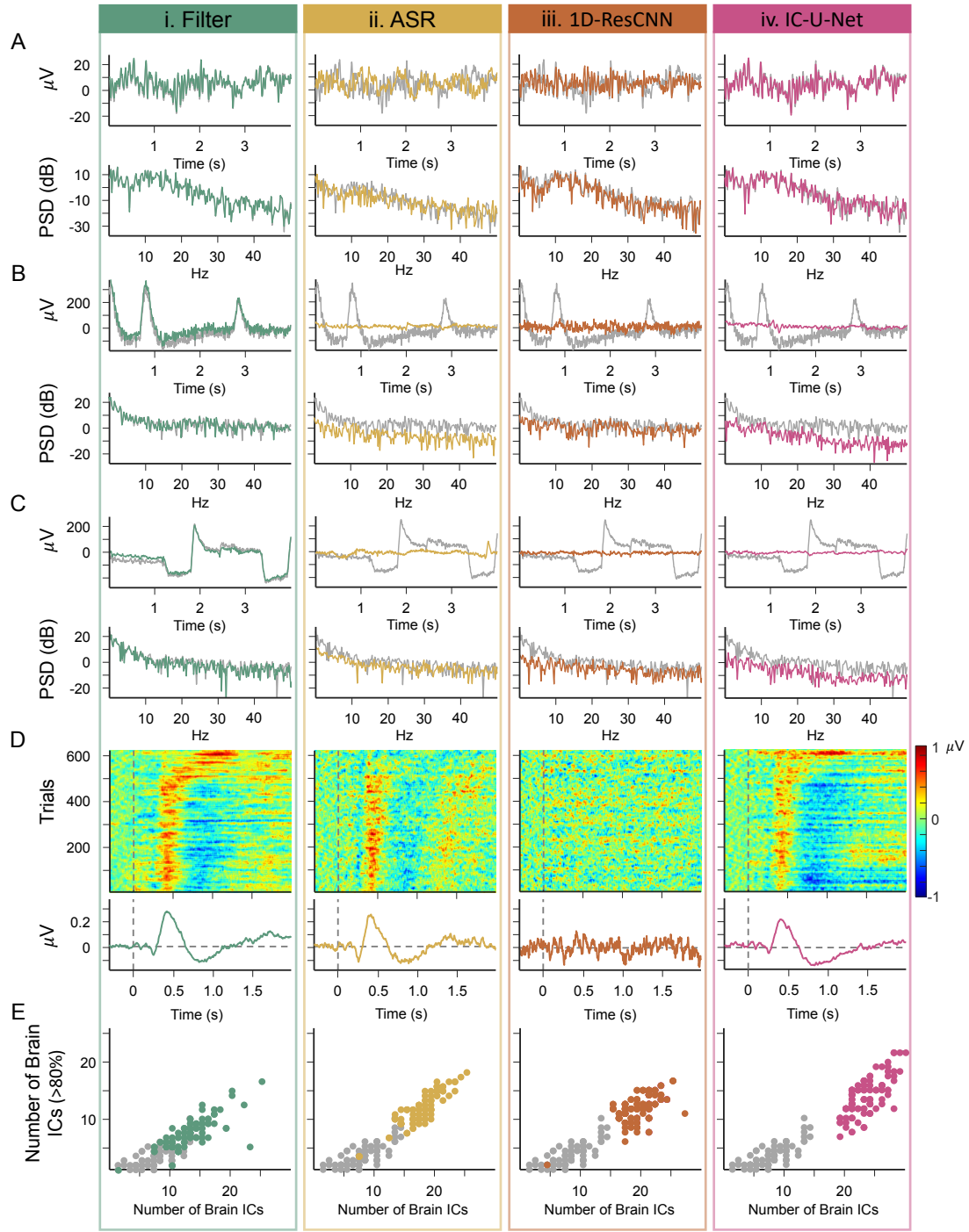

Figure 6. Performance comparison of the filter, ASR, 1D-ResCNN, and IC-U-Net models in correcting EEG signals obtained from a lane-keeping driving experiment. (A-C) Time series and PSDs of one clean EEG segment and two 4-s EEG segments contaminated by eye-blinking and square-like artifacts (gray traces), which were corrected by the aforementioned four artifact removal methods. (D) ERP image and average ERP of the Cz channel after artifact removal using different methods. The results in A and B were obtained using a specific lane-keeping dataset: s05_061019m. (E) Number of Brain ICs decomposed from 76 lane-keeping datasets; the x-axis indicates the number of Brain ICs, and the y-axis indicates the number of Brain ICs whose probability of being brain activity exceeds 80%.

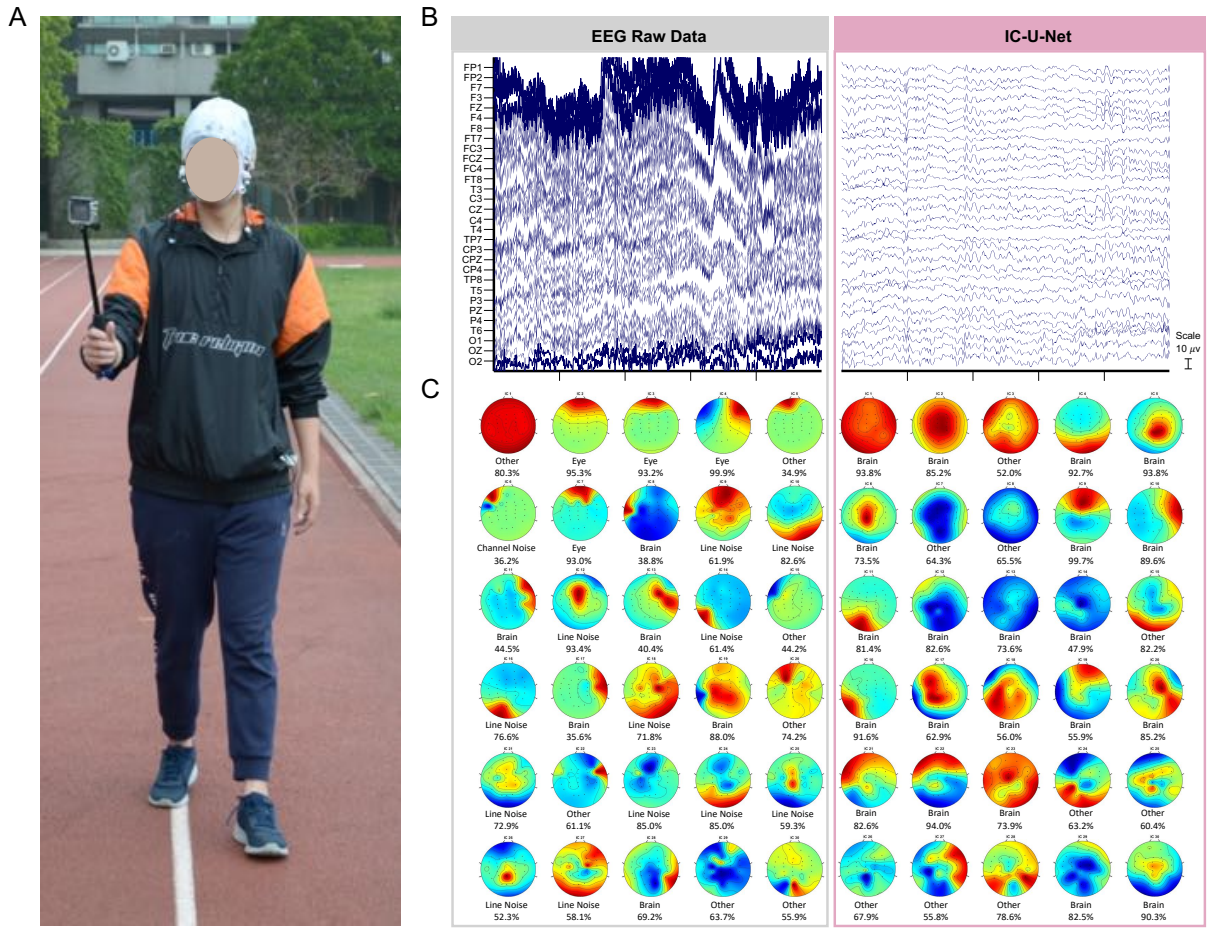

Figure 7. Using IC-U-Net to remove artifacts in EEG data collected during a walking experiment. (A) 32-channel EEG recorded from a subject who was walking along a running track in a university stadium. (B) Plots of raw (left) and reconstructed (right) time series signals. (C) ICs of raw (left) and reconstructed (right) signals decomposed by ICA and recognized by ICLabel. The numbers indicate the source class probability of each IC.